\documentstyle[art12,openbib]{article}
\begin{document}
\title{Relativistic  effect on low-energy nucleon-deuteron  scattering}
\author{Sadhan K. Adhikari and Lauro Tomio \\
Instituto de F\'\i sica Te\'orica, Universidade Estadual Paulista \\
01405-000 S\~{a}o Paulo, S\~{a}o Paulo, Brasil }
\vspace{1 true cm}
\date{}
\maketitle
\begin{abstract}
 The relativistic effect on differential cross sections, nucleon-to-nucleon
and nucleon-to-deuteron polarization transfer coefficients, and the spin
correlation function, of nucleon-deuteron elastic scattering is investigated
employing several three-dimensional relativistic three-body equations and
several nucleon-nucleon potentials.  The polarization transfer coefficients
are found to be sensitive to the details of the nucleon-nucleon potentials
and the relativistic dynamics employed, and prefer trinucleon models with the
correct triton binding energy.

(To appear in Physical Review C)
\end{abstract}

\section{INTRODUCTION}
Recently,  there have been many benchmark nonrelativistic calculations of the
three-nucleon system involving realistic two- and three-nucleon
potentials\cite{bench,bench2,ak}.  However, both the bound-state and
low-energy scattering calculations of this system involve large momentum
components of the wave function or the $t$ matrix elements, and this demands
for a relativistic treatment of the problem. Relativistic dynamical
calculations in the three-nucleon problem have been mainly restricted to the
study of the three-nucleon bound
state\cite{rupp1,rupp2,garci1,mach,carl,adhi1} with few exceptions where
relativistic effect on the nucleon-deuteron (nd) scattering length has been
studied\cite{adhi1,garci2}.  Both the four-dimensional Bethe-Salpeter-Faddeev
equation\cite{rupp1,rupp2} in some approximate form and several types of
three-dimensional reductions of this equation have been employed in numerical
calculations\cite{rupp1,rupp2,garci1,mach,adhi1,garci2}.

We study the effect of  relativistic dynamics on nd elastic scattering  by
employing several nucleon-nucleon potential models and four types of
three-dimensional relativistic scattering equations suggested
recently\cite{adhi2}. These  relativistic equations  obey conditions of
relativistic covariance and two- and three-particle unitarity. At present
time, one of the practical and feasible ways of performing a relativistic
three-nucleon calculation is through these three-dimensional equations.  At
this point it should be noted that the solution of the approximate BS
equation in ladder form, as has been performed recently\cite{rupp1,rupp2}, is
not necessarily a superior way of dealing with the relativistic
effect\cite{mach,gross}, partly because of the problems in interpreting the
four-dimensional potential in the BS equation, and partly because of the
difficulty in extracting a meaningful single-particle limit from the two
particle BS equation in the ladder form.

In our previous investigation\cite{adhi1} we studied the effect of
relativistic dynamics on the trinucleon binding energy and the  $S$-wave nd
scattering length.  Here we study the relativistic effect on differential
cross sections, various nucleon-to-nucleon and nucleon-to-deuteron
polarization transfer coefficients, and the spin correlation function,
$C_{xx}$ of nd  elastic scattering.  The present  study indicates that these
polarization transfer coefficients are very sensitive to the details of the
nucleon-nucleon potentials  and the dynamics employed in  three-nucleon
calculation, and favor models with correct triton binding energy.

The sensitivity of the nucleon-to-nucleon polarization transfer coefficients
of nd elastic scattering, specially $K^{y'}_{y}$, at low energies ($<$ 10
MeV) to the off-shell behavior of the nucleon-nucleon potential has been
known for quite some time\cite{ple}.  Recently, in relation to a study of the
three-nucleon problem with realistic meson-theoretic potentials it has been
observed\cite{bench,glo} that these polarization transfer coefficients at
22.7 MeV are very sensitive to the tensor part of the nucleon-nucleon
potential. It was found that the theoretical calculation employing the Bonn A
meson-theoretic potential reproduced the results of nucleon-to-nucleon
polarization transfer coefficient $K^{y'}_{y}$ better than the calculation
based on the meson-theoretic Paris potential. As the tensor parts of these
two potentials are very different, it was concluded\cite{bench,glo} that this
finding supports the weak tensor force of the Bonn A  potential  as being
more realistic than the stronger tensor force of the Paris potential.

A previous study\cite{adhi3} demonstrated that three-nucleon models based on
central nucleon-nucleon potentials can reproduce nucleon-to-nucleon
polarization transfer coefficient $K^{y'}_{y}$ quite well, provided that
these models also reproduce the correct triton binding energy. In view of
this it is unlikely that the nucleon-to-nucleon polarization transfer
coefficient $K^{y'}_{y}$ should carry much new information about tensor
nucleon-nucleon potentials which is not implicit  in the results of the
triton binding energy.   The Bonn A tensor potential reproduces for triton
binding energy 8.38 MeV, whereas the Paris potential yields 7.47 MeV. As the
Bonn A potential reproduces the experimental triton binding energy (8.48 MeV)
better than the Paris potential, it also reproduces the above $K^{y'}_{y}$
better than the Paris potential.  This casts doubt on the conclusion about
the superiority of the Bonn A tensor potential over that of the Paris
potential\cite{bench,glo}. In our previous study\cite{adhi3} only
nonrelativistic three-nucleon models were used. In this study we would like
to see if the inclusion of relativistic dynamics changes the general
conclusions of Ref.
\cite{adhi3}.

Though the magnitudes of relativistic corrections to the triton binding
energy, $B_t$,  and the $S$-wave spin-doublet nd scattering length, $a_{nd}$,
as emphasized in previous studies, are interesting, it is most relevant to
see if meaningful physics could be extracted from the relativistic treatment
of the three-nucleon system. The nonrelativistic potential model calculations
of these two observables employing meson-theoretic nucleon-nucleon
potentials\cite{meson} did not allow us to extract meaningful informations
about the two- and three-nucleon interactions because of the correlated
behavior of the observables directly sensitive to these
interactions\cite{ak}. The  question to ask at this stage is whether the
relativistic treatment of the three-nucleon problem  is expected to change
the scenario.

We present the present model in Sec. II. In Sec. III we present numerical
results. Finally, in Sec. IV a brief summary of the conclusions are given.

\section{Nucleon-deuteron models}

The models that we shall use have been developed recently\cite{adhi1,adhi2}.
Here we  present a brief summary of these models.

The relativistic two-nucleon dynamics for a central $S$ wave potential is
taken to be governed by the following partial-wave Blankenbecler-Sugar (BlS)
equation\cite{bls}
\begin{equation}
t(q',q,k^2)=V(q',q)+4{\pi}\int_0^\infty p^2 dp  \frac{m}{\omega_p}
V(q',p)\frac{1} {k^2-p^2+i0}t(p,q,k^2),
\label{B} \end{equation}
where  $\omega_p =(m^2+p^2)^{1/2}$.  The nonrelativistic two-nucleon dynamics
is taken to be governed by the Lippmann-Schwinger (LS) equation.

We take the relativistic nucleon-nucleon potential in  the following form
\begin{equation}
  [V_n(q',q)]_{rel} = -\lambda _n [v_n(q')]_{rel} [v_n(q)]_{rel},
\label{1} \end{equation}
where $n=0$ (1) represents the spin triplet (singlet) state, and the
subscript  $rel$ ($nr$) denotes relativistic (nonrelativistic).

The relativistic $t$ matrix in this case at the square of the center of mass
(c.m.)  energy $s=4(m^2+k^2)$ is given by
\begin{equation}
[t_n (q',q,k^2)]_{rel} = [v_n(q')]_{rel} [\tau_n^{-1}(k^2)]_{rel}
[v_n(q)]_{rel},
\label{3} \end{equation}
where
\begin{equation}
[\tau_n (k^2)]_{rel} = -\frac{1}{\lambda _n} - 4{\pi}\int^{\infty}_0 q^2 dq
\left( \frac{m}{\omega_q}\right) \frac {[v_n(q)]^2_{rel} }{k^2-q^2+i0}.
\label{4} \end{equation}
 We  generate a nonrelativistic  two-nucleon $t$ matrix, phase-equivalent to
its relativistic version by the following transformation for the form-factors
\begin{equation}
[v_n(q)]_{nr} = (\sqrt {m/\omega_q}) [v_n(q)]_{rel},
\label{13} \end{equation}
 so that \begin{equation} [t_n(q',q,s)]_{nr} =
[v_n(q')]_{nr}[\tau_n^{-1}(k^2)]_{rel}[v_n(q)]_{nr}, \label{14}
\end{equation} The functional form of $[\tau]_{rel}$ of Eq. (\ref{14}) is
exactly identical to its relativistic counterpart (\ref{4}).  This procedure
generates phase-equivalent two-nucleon potentials to be used in
nonrelativistic and relativistic  three-nucleon problem.

The three-nucleon problem is solved with the following one-nucleon-exchange
three-nucleon Born term\cite{adhi1}
\begin{equation}
B_{n,n'}(\vec p, \vec q) =  {v_n({\cal P})v_{n'}({\cal Q})} G(\vec p,\vec q).
\label{101} \end{equation}
In the nonrelativistic case the propagator $G(\vec p, \vec q)$ is  given by,
\begin{equation}
G_{nr}(\vec p,\vec q,E) =(p^2+q^2+pqx -mE -i0)^{-1},
\label{7} \end{equation}
with  ${\cal P}^2 = p^2/4 +q^2 +pqx$,   and $ {\cal Q}^2 = q^2/4 +p^2 +pqx$,
where $x$ is the cosine of the angle between $\vec p$ and $\vec q$.

 In  the relativistic case we use the following propagators in  Eq.
(\ref{101}) \cite{adhi1,adhi2} \begin{equation} G_{{\cal A}}(\vec p,\vec q,s)
= \frac {2(\omega_p+\omega_q+\omega_{pq})} {\omega_{pq}
[(\omega_p+\omega_q+\omega_{pq})^2 - s-i0]}; \label{23} \end{equation}
\begin{equation} G_{{\cal B}}(\vec p,\vec q,s) = \frac {2(\omega_p+\omega_q)}
{\omega_{pq} [(\omega_p+\omega_q)^2 - (\sqrt s -\omega_{pq})^2-i0]};
\label{24} \end{equation} \begin{equation} G_{{\cal C}}(\vec p,\vec q,s) =
\frac {1} {\omega_{pq} [\omega_p+\omega_q+\omega_{pq} - \sqrt s-i0]};
\label{25} \end{equation} \begin{equation} G_{{\cal D}}(\vec p,\vec q,s) =
\frac {2 (\omega_q+\omega_{pq})} {\omega_{pq} [(\omega_q+\omega_{pq})^2 -
(\sqrt s - \omega_p)^2-i0]}, \label{26} \end{equation} with $  {\cal P}^2 =
(\omega _q+\omega_{pq})^2/4-p^2/4 -m^2,$ and $  {\cal Q}^2 = (\omega
_p+\omega_{pq})^2/4-q^2/4 -m^2$.  Here we use notations $\omega_p
=(m^2+p^2)^{1/2}$, $\omega_{pq} =[m^2+(\vec p +\vec q)^2]^{1/2}$, etc.  The
spin variables are treated nonrelativistically  in all cases.

 In Eqs. (\ref{23}) - (\ref{26}) the parameter $s$ is the square of the total
c.m. energy of the three-particle system. All these propagators satisfy
conditions of relativistic three-particle unitarity, governed by that part of
the denominator in these propagators which corresponds to the pole for
three-particle propagation in the intermediate state, e.g., at $\sqrt s =
\omega_p+\omega_q+\omega_{pq}$.  The condition of relativistic three-particle
unitarity in these propagators is manifested in having the same residue at
this pole. These equations also  satisfy conditions of two-particle
unitarity.

 Equation (\ref{23}) was  advocated by Aaron, Amado, and Young\cite{aaron}
and obeys time-reversal symmetry, e.g. $G(\vec p,\vec q,s) = G(\vec q,\vec
p,s)$.  Equations (\ref{24}) and (\ref{25}) also have this virtue of Eq.
(\ref{23}).  The propagator $G_{{\cal C}}$ was suggested long
ago\cite{ahmad}. It has been shown\cite{adhi4} that the propagator $G_{{\cal
D}}$ follows from a suggestion by Ahmadzadeh and Tjon\cite{ahmad}. However,
the propagator $G_{{\cal D}}$ has never appeared in this form before.
Previous numerical applications\cite{rupp1,rupp2,garci1,mach}  of this
propagator used unnecessary approximations which violated conditions of
unitarity and covariance\cite{adhi4}.  Physically, these propagators differ
in the way the particle and antiparticle contributions appears in the kernel
of the integral equation. In the propagator $_{{\cal C}}$, for example, there
is no antiparticle contribution.

\section{Numerical Results}

For  two-nucleon separable potentials in  spin-triplet and spin-singlet
channels we take the following Yamaguchi and Tabakin form-factors\cite{yt},
recently used by Rupp and Tjon\cite{rupp2}:
\begin{equation}
g_Y(q)=\frac {1} {q^2+\beta ^2},
\label{30} \end{equation}
\begin{equation}
g_T(q)=\frac{q^2+\nu ^2}{q^2+\gamma ^2} \times
\frac{q_c^2-q^2}{(q^2+\beta ^2)^{\kappa}}, \kappa=1.5, 2.
\label{31} \end{equation}
The Yamaguchi potential will be referred to as Y, and the Tabakin potential
with $\kappa = 1.5, 2$ will be referred to as T-1.5 and T-2, respectively.
The constants of these potentials  are given in Ref.  \cite{adhi1}.

 Tabakin-type nucleon-nucleon potentials yield nucleon-nucleon phase shifts
in better agreement with experiment, which change sign at higher energies,
compared to the Yamaguchi potential. If Tabakin-type potential is used in
both $^3S_1$ and $^1S_0$ spin channels, it leads to a triton  ground state of
several hundred MeV's\cite{bsc}.  However, we shall  use  the Tabakin
potential in one of the nucleon-nucleon spin channels and Yamaguchi in the
other and this   does not lead to a collapsed triton and leads to trinucleon
observables in better agreement with experiment and realistic calculations.
The numerical calculation is also simplified by an order of magnitude in this
model. We perform this  `cheap' study with a view to conclude if a more
realistic calculation is worth the price.

At positive energies the three-particle equations were solved by the
technique of contour rotation\cite{ak}. In the relativistic case, as was
already noted before\cite{rupp1,rupp2} in the bound-state problem, more care
was needed to obtain a converged result in the scattering calculation. Some
48 mesh points were needed for obtaining  converged relativistic  scattering
results with a rotation angle of 5 degrees, whereas some 24 mesh points were
enough to obtain the nonrelativistic scattering results to the same degree of
precision.

We calculated  the $S$ wave spin-doublet nd scattering length, $a_{nd}$,
differential cross section, spin correlation function, nucleon-to-nucleon and
nucleon-to-deuteron polarization transfer coefficients of nucleon-deuteron
elastic scattering and the triton binding, $B_t$ in the nonrelativistic case
as well as with each of the four versions of  relativistic formulations
${{\cal A}}-{{\cal D}}$.  Propagator ${{\cal A}}$ has been used before in
numerical calculations of the three-nucleon
problem\cite{rupp1,rupp2,garci1,garci2}. To the best of our knowledge,
propagators ${{\cal B}}$, ${{\cal C}}$, and ${{\cal D}}$ are new and have
never been used before in  the three-nucleon problem.

The results for triton binding energies and the $S$ wave doublet scattering
lengths in the present models have recently appeared\cite{adhi1}.  All the
relativistic propagators increase the triton binding energy, $B_t$,  in
relation to the nonrelativistic case, except propagator ${{\cal C}}$ which
reduces the binding.  This tendency, also observed  in previous
calculations\cite{rupp1,rupp2,carl}, has been justified recently by
theoretical arguments\cite{adhi1}. The relativistic correction to $B_t$
varies from $-$0.3 MeV to 0.7 MeV in different situations.  We observed in
numerical calculations the following general inequality
\begin{equation}
 (B_t)_{{\cal D}}, (B_t)_{{\cal B}} > (B_t)_{{\cal A}} > (B_t)_{nr} >
(B_t)_{{\cal C}}.
\end{equation}
These results are summarized in Fig. 1 where we  plot $B_t$ versus $a_{nd}$
for the present nonrelativistic and relativistic model calculations, as well
as for many other nonrelativistic calculations taken from the literature
\cite{opt,ad}. In nonrelativistic calculations a correlation was observed
between $B_t$ and $a_{nd}$\cite{ak,phillips} as a result of on- and off-
shell variations of two- and three-nucleon potentials. The relativistic
calculations of Fig. 1 differ in employing different relativistic dynamics
and nucleon-nucleon potentials. The trend of the relativistic calculations is
identical to that of the nonrelativistic calculations.  Hence, the effect of
including relativistic dynamics in the three-nucleon problem can not be
distinguished from the effect of varying the two- and three-nucleon
potentials in  nonrelativistic calculations.

Next we present results for some of the low-energy nd scattering observables.
We chose  to exhibit five models for nd elastic scattering covering a wide
range of variation of triton binding energy. They are

Model A: nonrelativistic dynamics, triplet Y singlet Y, $B_t$=10.65 MeV,

Model B: relativistic propagator  ${{\cal B}}$,  triplet T-2 singlet Y,
$B_t$=8.34    MeV,

Model C: relativistic propagator ${{\cal A}}$,  triplet T-2 singlet Y,
$B_t$=8.14    MeV,

Model D: relativistic propagator ${{\cal C}}$,   triplet T-2 singlet Y,
$B_t$=7.91    MeV,

Model E: nonrelativistic dynamics,  triplet Y singlet T-2, $B_t$=7.69    MeV.

Three of these models are relativistic and two nonrelativistic.  Note that
the models A, B, C, D, and E produce triton binding energy in monotonically
decreasing order. We performed calculations for other combinations of
nucleon-nucleon potentials and propagators which follow the general trend of
results obtained with these five illustrative models.

In Fig. 2 we plot the elastic differential scattering cross section at
nucleon laboratory energies, $E_n$,  of 10,  22.7, and 100 MeV. At lower
energies the relativistic effect on this observable is small and possibly
could be ignored. At higher energies (100 MeV) this effect is small at
forward and backward angles, but could be reasonable near the minimum of the
cross section. The nd elastic differential  cross section is mainly dominated
by the spin quartet state and no specific correlation of the cross section
with the triton binding energy was observed. In the following we consider
several scattering observables which are correlated with the triton binding
energy.

It is most relevant to consider the nucleon-to-nucleon polarization transfer
coefficient, $K^{y'}_{y}$ of nd elastic scattering.  In Refs.\cite{ple} it
has been claimed that this observable at $E_n$ = 10 MeV is very sensitive to
the off-shell behavior of the nucleon-nucleon $S$ wave interaction. A
calculation based on Yamaguchi nucleon-nucleon potential did not reproduce
the experimental results for this observable whereas that based on a
meson-theoretic nucleon-nucleon potential\cite{meson} could explain the
experimental results.  The correct off-shell behavior of the meson-theoretic
potential was made responsible for this\cite{ple}. More recently, it has been
claimed\cite{bench,glo} that this observable at $E_n$ = 22.7 MeV is very
sensitive to the tensor force of the nucleon-nucleon interaction.  A
calculation using the  Paris nucleon-nucleon potential could not reproduce
the experimental results for this observable whereas that using the Bonn A
potential could satisfactorily account for  the experimental results.  The
`correct' tensor force of the Bonn A potential has been made responsible for
this\cite{bench,glo}. In Ref. \cite{adhi3} it was pointed out that
$K^{y'}_{y}$ is correlated with the triton binding energy $B_t$ (or the spin
doublet nd scattering length $a_{nd}$) in a dynamical calculation and the
experimental data for $K^{y'}_{y}$ favors a  three-nucleon model with the
correct triton binding energy.  In the study of Ref.  \cite{ple} the
meson-theoretic nucleon-nucleon potential yielded a $B_t$, and hence
$K^{y'}_{y}$, closer to experiment than the Yamaguchi nucleon-nucleon
potential. In Ref.  \cite{bench,glo} the Bonn A potential yields a $B_t$, and
hence $K^{y'}_{y}$, closer to experiment than the Paris nucleon-nucleon
potential. Let us see if the inclusion of the relativistic dynamics changes
the above scenario.

In Fig. 3 we plot nucleon-to-nucleon polarization transfer coefficient
$K^{y'}_{y}$ of nd elastic scattering at $E_n$ = 10 MeV for the nd models A,
B, C, D, and E.  The results at 22.7 MeV for this observable are shown in Fig
4. We find that $K^{y'}_{y}$ is sensitive to both the nucleon-nucleon
potential models as well as to the dynamics employed. However, the minimum of
$K^{y'}_{y}$ at about $\theta_{c.m.} = 110$ degrees is found to be correlated
with the triton binding energy as in the nonrelativistic case\cite{adhi3}.
In view of the present result and that of Ref.  \cite{adhi3} it is highly
likely that the better agreement of the Bonn A potential calculations for
$K^{y'}_{y}$ with  the experiment over that of the Paris potential
calculation at $E_n$ = 22.7 MeV is due to more precise $B_t$ produced by the
former potential.  Also, as  simple central potential models of nd scattering
could reproduce the results for $K^{y'}_{y}$ reasonably well, it seems
unlikely that it will really provide new information about the tensor
nucleon-nucleon potential besides those already contained in the value of
triton binding energy. At higher energies, for example, at $E_n$ = 100 MeV,
the sensitivity of $K^{y'}_{y}$ to the dynamics is found to be highly
reduced; all five models yield essentially the same result for $K^{y'}_{y}$
independent of triton binding.

Next we exhibit in Fig. 5 the spin correlation function $C_{xx}$ at 10 MeV
for the five nd potential models. Again the results for $C_{xx}$ are
sensitive to the nd models, and arround $\theta_{c.m}$ = 110 degrees the
results follow the order of the triton binding energy. The experimental
points in this case are taken from Ref. \cite{arvi}. In this case the
dispersion between the curves of the different models is mostly determined by
the corresponding triton binding energy and independent of the
nucleon-nucleon potential models and the three-particle propagators.

Recently, it has been observed\cite{glo1} that the nucleon-to-deuteron
polarization transfer coefficients of nd elastic scattering are also
sensitive to nd models employed. In order to study this sensitivity we plot
in Fig. 6 and 7 nucleon-to-deuteron coefficients $K^{y'}_{y}$ and
$K^{x'}_{x}$ at $E_n$ = 10 MeV for nd elastic scattering, respectively.  The
results for different models are separated again according to the value of
triton binding energy.

It has been pointed out in Ref.  \cite{adhi3} that though a $S$ wave
separable potential model give a good description of the polarization
transfer coefficients of nd elastic scattering, tensor and higher partial
waves of nucleon-nucleon potentials are needed for their accurate
description. For an $S$ wave model one should have for the nucleon-to-nucleon
polarization transfer coefficients
\begin{equation}
K^{x'}_x = -K^{y'}_y sin\theta_{lab} = K^{y'}_y cos\theta_{lab}.
\end{equation}
The differences
\begin{equation}
\Delta K= -K^{x'}_x  -K^{y'}_y sin\theta_{lab}
\end{equation}
and
\begin{equation}
\Delta K'=-K^{x'}_x + K^{y'}_y cos\theta_{lab},
\end{equation}
are good measures of the effect of noncentrality of the nucleon-nucleon
potental. However, to quantize such effects from a study of $\Delta K$ and
$\Delta K'$  high precision experimental results are needed. As the functions
$\Delta K$ and $\Delta K'$ are supposed to carry informations about the
tensor-force and higher partial waves of nucleon-nucleon interaction,
experimentalists are encouraged to provide accurate data for these
observables. In the absence of these interactions both $\Delta K$ and $\Delta
K'$ are identically zero. Numerical calculations employing meson-theoretic
nucleon-nucleon potentials are essential for an accurate description of the
polarization transfer coefficients. However, as the experimental polarization
transfer coefficients have large error bars, both the differences $\Delta K$
and $\Delta K'$ are small with even larger errors\cite{adhi3}. Consequently,
an $S$-wave treatment of the problem for drawing general conclusions as has
been done here is justified.

Though we exhibit here in the figures results for five specific models, we
performed and studied the numerical calculations for several more models. The
general trend found in the case of these five models was observed in all
cases.  It is well known that  the present $S$ wave separable potential
models do not give a good description of the nd scattering at low energies.
In spite of this,  we have drawn some general conclusions which should be
valid in realistic situations. We summarize these conclusions in the next
section.

\section{Conclusions}

In conclusion, this is the first systematic study  of  relativistic effect on
the trinucleon bound state and scattering employing  combinations of
Yamaguchi and Tabakin type potentials for the singlet and triplet
nucleon-nucleon channels and four types of relativistic three-particle
scattering equations.

We find in the calculations employing relativistic dynamics that $B_t$ is
correlated to $a_{nd}$ as in nonrelativistic model calculations with
variation of nucleon-nucleon potentials on- and off-shell.  Hence it will be
difficult to separate the effect of such variation of potentials from that of
introducing the relativistic dynamics.  This confirms the existence of a
shape-independent approximation to these observables even after inclusion of
the relativistic effect\cite{opt}.

We observe that the nucleon-to-nucleon and nucleon-to-deuteron polarization
transfer coefficients of nd elastic scattering are very sensitive to the
details of potential model and relativistic dynamics. This sensitivity should
be  highly reduced once differnent models yield the same triton binding
energy. In view of this it is highly improbable that the results for
nucleon-to nucleon polarization transfer coefficient, $K^{y'}_y$, favors the
weak tensor force of the Bonn A potential as has been concluded by Clajus et
al. recently\cite{glo}.  From a study of the nd elastic differential cross
section,  mainly dominated by the spin quartet state,  no specific
correlation with the triton binding energy was observed. However, the
low-energy spin-correlation function is found to be correlated with the
triton binding energy.

Of course, there are  other observables for the three-nucleon system, which
should be directly sensitive to relativistic effect, such as the charge form
factors. Because of the presence of the possible large effect of
meson-exchange currents and of the non-nucleonic components in the nucleus,
such observables are not easily tractable, and it has so far been difficult
to draw model independent conclusion from studies of these
observables\cite{bench,bench2}.

We are aware that there is an inherent flexibility in deciding on the
relativistic dynamics, in treating the spin variables relativistically, and
in deciding the correct form of two- and three-nucleon potentials. We are far
from exhausting all possibilities. But the tendency of existing the
shape-independent approximation is so strong that we do not believe the
present conclusions to be so peculiar as to be of no general validity. Hence
a relativistic framework may reduce the still existing discrepancy between
theory and experiment, but this may not enhance our knowledge of the
three-nucleon system.

We thank Dr. T. Frederico for stimulating discussions.  LT thanks the
European Centre for Theoretical Studies in Nuclear Physics and Related Areas
(ECT) at Trento, Italy for its hospitality and local support.  The work is
supported in part by the Conselho Nacional de Desenvolvimento - Cient\'\i
fico e Tecnol\'ogico (CNPq) of Brasil.

\vskip 1cm

{\bf Figure Caption}
\begin{itemize}
\item[1.] The $B_t$ versus $a_{nd}$ plot for various trinucleon models: the
present relativistic models ($\diamond$), the present  nonrelativistic models
($+$), and other nonrelativistic models taken from the literature
($\times$)\cite{ad}.
\item[2.] The nd elastic differential scattering cross section at different
incident neucleon energies.
\item[3.] The nucleon-to-nucleon polarization transfer coefficient $K^{y'}_y$
of nd elastic scattering for $E_n=10$ MeV. The curves labelled A, B, and E
correspond to models A, B, and E, respectively. The two curves between B and
E correspond to models C and D in the order of decreasing triton binding
energy.  The dispersion between the curves is in the order of the variation
of the triton binding energy.
\item[4.] Same as Fig. 3 for $E_n=22.7$ MeV. The curve for model E almost
coincides with that of model D in this case and is not shown in this case.
\item[5.] The spin correlation function of nd elastic scattering for $E_n=10$
MeV.  For other details see Caption of Fig. 3.
\item[6.] The nucleon-to-deuteron polarization transfer coefficient
$K^{y'}_y$ of nd elastic scattering for $E_n=10$ MeV. For other details see
Caption of Fig. 3.
\item[7.] The nucleon-to-deuteron polarization transfer coefficient
$K^{x'}_x$ of nd elastic scattering for $E_n=10$ MeV.  For other details see
Caption of Fig. 3.
\end{itemize}
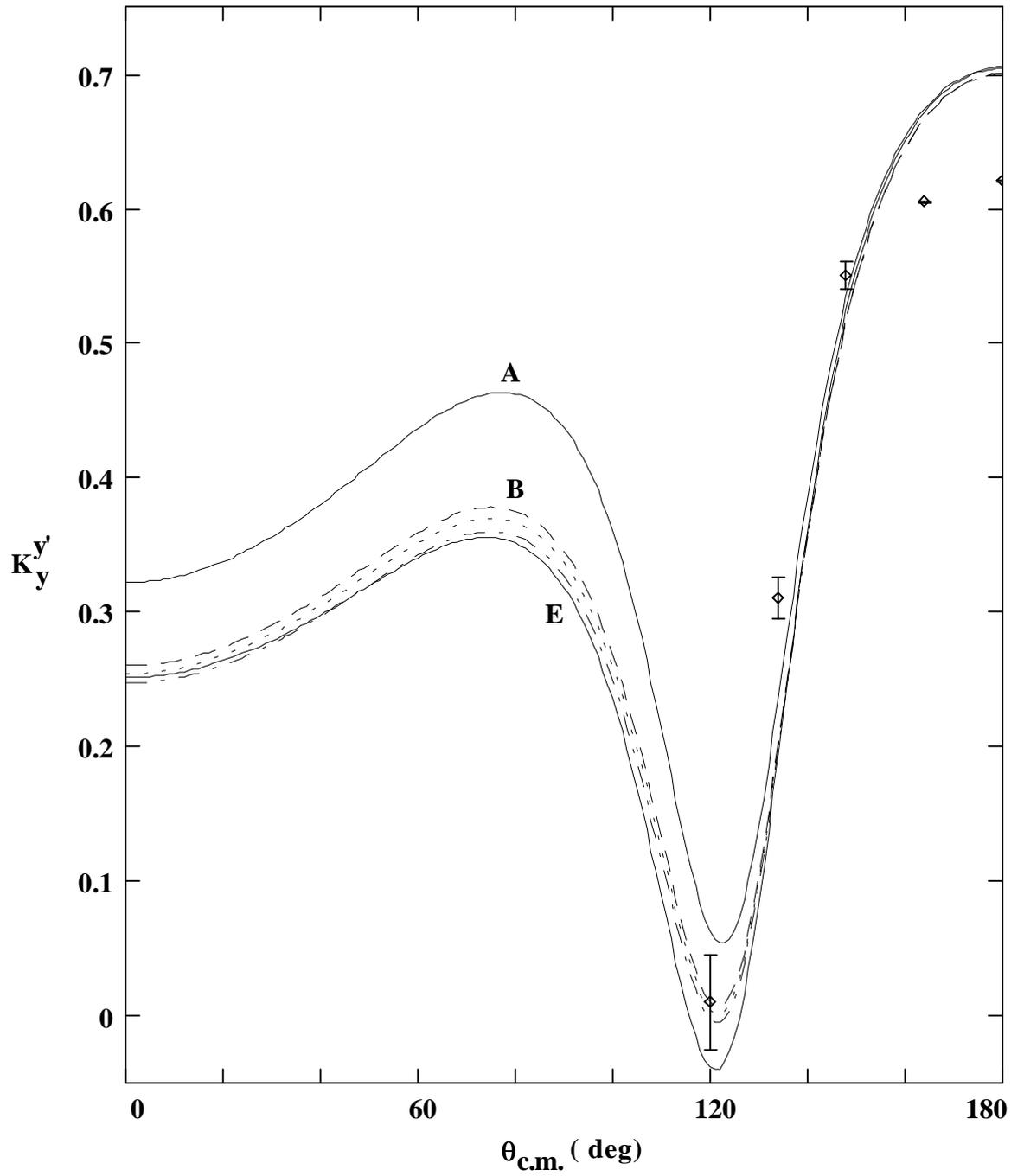
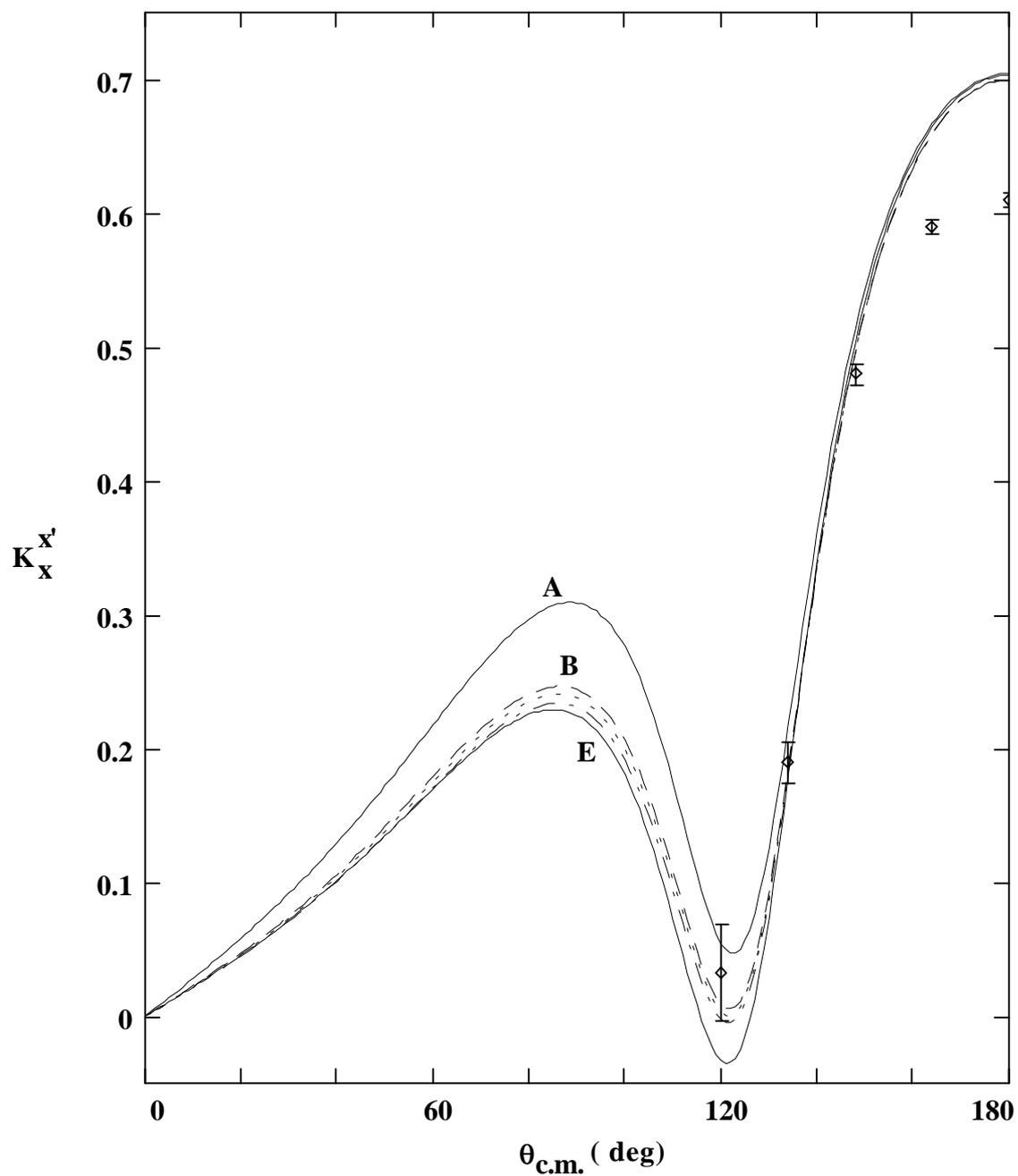
\end{document}